\documentclass{appolb}
\usepackage{epsfig}
\def\be{\begin{equation}}
\def\ee{\end{equation}}
\def\bea{\begin{eqnarray}}
\def\eea{\end{eqnarray}}

\def\d{{\mbox{\rm d}}}

\begin{document}
\title{Bose-Einstein or HBT correlations \\ and the anomalous dimension of QCD%
\thanks{Presented by T. Cs\"org\H{o} at the XXXIV International
Symposium on Multiparticle Dynamics, Sonoma County, California, USA, 
July 26 - August 1, 2004}%
}
\author{T. Cs\"org\H{o}, S. Hegyi, 
\address{MTA KFKI RMKI, H - 1525 Budapest 114, P.O.Box 49, Hungary}
\and
T. Nov\'ak 
\address{University of Nijmegen, NL - 6525 ED Nijmegen, Toernooiveld 1}
\and
W. A. Zajc
\address{Department of Physics, Columbia University, 538 W 120th Street, New York, NY 10027, USA}
}
\maketitle
\begin{abstract}
The Bose-Einstein (or HBT) correlation functions are evaluated
for the fractal structure of QCD  jets.  These correlation functions 
have a stretched exponential (or L\'evy-stable) form. 
The anomalous dimension of QCD determines  the L\'evy index of 
stability, thus the running coupling constant of QCD 
becomes measurable with the help of two-particle Bose-Einstein correlation functions. 
These considerations are tested on NA22 and UA1 two-pion correlation data.
\end{abstract}
  
\section{Introduction}
	The study of fractal phenomena was introduced to high energy particle and nuclear physics by
	Bialas and Peschanski in ref.~\cite{Bialas:1985jb},  see also ref.\cite{Bialas:1990gt} for
	a review.  In QCD, jets emit jets that emit additional jets and so on.
	The resulting fractal structure of QCD jets was explored with the
	help of a beautiful geometric picture in refs.
	\cite{Dahlqvist:1989yc,Gustafson:1990qi,Gustafson:1990qk}.
	These ideas were developed further by the Lund group in refs.  
	~\cite{Gustafson:1991ru,Gustafson:1992uh,Andersson:1995jv},  by
	Dokshitzer and Dremin in ref.~\cite{Dokshitzer:1992df} as well as by
	Ochs and Woisek in refs.~\cite{Ochs:1992tg,Ochs:1994rt}.
	Both theoretical and experimental aspects of the so-called intermittency or 
	fractal structures in high energy physics were reviewed by 
	De~Wolf, Dremin and Kittel in ref.~\cite{DeWolf:1995pc}.

	Bialas realized, that  Bose-Einstein correlations and 
	intermittency might be deeply connected~\cite{Bialas:1992ca}. 
	The mathematical properties of Bose-Einstein correlation functions 
	for L\'evy stable (convolution invariant) sources were written up by three of us in 
	refs.~\cite{Csorgo:2003uv,Csorgo:2004ch}.
	Here we add a physical interpretation and show, 
	that the fractal properties of QCD cascades can naturally  be measured
	by the L\'evy index of stability of Bose-Einstein correlations. 
	Our analytical results are similar in spirit to the numerical investigations
	of Wilk and collaborators in ref.~\cite{Utyuzh:1999zg}.

\section{(Multi)fractal structure of the QCD jets }
\label{s:jets}
	In this section we recapitulate earlier theoretical results of the Lund group,~
	\cite{Dahlqvist:1989yc,Gustafson:1990qi,Gustafson:1990qk,Gustafson:1991ru,Gustafson:1992uh,Andersson:1995jv},
	that related the properties of QCD cascades to intermittency. These results are
	based on a beautiful geometric interpretation of the color dipole 
	picture, and on an infrared stable measure on parton states related
	to hadronic multiplicity.

\begin{figure}[!thb]
\begin{center}
\vspace{-0.5cm}
\includegraphics[angle=0,width=\textwidth]{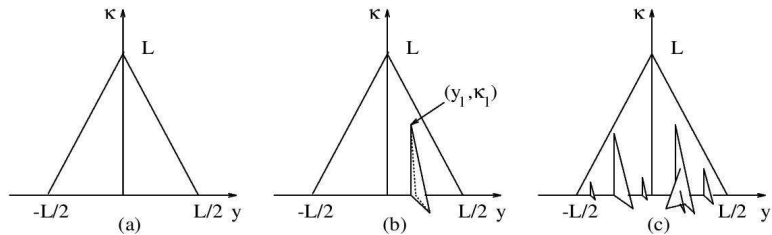}
\vspace{-1.0cm}
\end{center}
\caption[*]{
\label{fig:fractal-jets}{\small The phase-space of QCD jets in the $(y,\kappa)$ 
 	plane, where $\kappa = \log(k_t^2)$.
	(a) The phase-space available for a gluon emitted by a high energy $q\overline{q}$
	system is a triangular region in the $(y,\kappa)$ plane. 
	(b) If one gluon is emitted at $(y_1,\kappa_1)$,
   	the phase-space for a second (softer) gluon is
	given by the area of this folded surface. 
	(c) The total gluonic phase-space can be described by this
	multifaceted fractal surface~\cite{Dahlqvist:1989yc,Gustafson:1990qi,Gustafson:1990qk}.
} 
}
\end{figure}
	
	A high energy $q\overline{q}$ system radiates gluons according to the dipole formula    
\begin{equation}
  dn=\frac{3 \alpha_s}{4\pi^2} \frac{ d k_\perp^2}{ k_\perp^2} dy d\phi,
\end{equation}
	hence the phase-space for the emission of a gluon is given by the relation
\begin{equation}
	| y| \le \frac{1}{2} \ln(s/k_\perp^2),
\end{equation}
	 which corresponds to the triangular region in a $(y,\ln k_\perp^2)$ diagram
 	as shown in Fig. 1 (a). 
	If two gluons are emitted, then the distribution of the hardest gluon
 	is described by eq. (1). The distribution of the second, softer, gluon corresponds to
 	two dipoles, the first is stretched between the quark and the first gluon, and 
 	the second between the first gluon and the anti-quark. 
 	The phase-space available for the second gluon corresponds to the
 	folded surface in Fig. 1 (b), with the constraint $k_{\perp,2}^2 < k_{\perp,1}^2$,
 	as the first gluon is assumed to be the hardest one.
	This procedure can be generalized so that the emission of a third, still softer
	gluon corresponds to radiation from three color dipoles, with $n$ gluons emitted already
       	the emission of the $(n+1)$-th gluon is given by a chain of $n+1$ dipoles. 
	Thus, with many gluons, the gluonic phase-space can be represented by a 
	multi-faceted surface as illustrated in Fig. 1 (c). 
	Each gluon adds a fold to the surface, which increases the phase-space for softer gluons.
	(Note, that in this process the recoils are neglected, 
	as is normal in leading log approximation).
	Due to its iterative nature, the process generates 
	a Koch-type fractal curve at the base-line.
	The length of this base-line of the partonic structure on Figure 1.c 
	is proportional to the particle multiplicity. 
	This curve is longer, when studied with higher resolution: 
	it is a fractal curve, embedded into the four-dimensional
	energy-momentum space, characterized by the fractal dimension
\be 
    d_f = 1 + \sqrt{\frac{3 \alpha_s}{2 \pi}}, 
\ee
	or one plus the anomalous dimension 
	of QCD~\cite{Dahlqvist:1989yc,Gustafson:1990qi,Gustafson:1990qk}.

    With the help of the Lund string fragmentation picture, this fractal in momentum space is mapped
    into a fractal in coordinate space,  and the constant of conversion is
    the hadronic string tension, $\kappa \approx 1$ GeV/fm. This
    mapping does not change the fractal properties of the curve.
    The emission of  softer and softer  gluons corresponds to
    a smaller and smaller modification of this curve,
    as a gluon with a very small transverse mass creates 
    a very small kink on the Lund string. Hence this process is infrared stable.

\section{\label{s:BEC} Bose-Einstein correlations for L\'evy stable source distributions}
	Let us discuss here the stability of the
	particle emitting source in QCD, and consider the Bose-Einstein correlation 
	functions for such sources.  The two-particle Bose-Einstein correlation function 
	is defined as the ratio of the two-particle invariant
	momentum distribution to the product of the single-particle invariant momentum
	distributions:
\begin{equation}
C_2({\mathbf k}_1,{\mathbf k}_2) = 
\frac{N_2({\mathbf k}_1,{\mathbf k}_2)}
{N_1({\mathbf k}_1)\, N_1({\mathbf k}_2)}.
	\label{e:cdef}
\end{equation}
	If long-range correlations can be neglected or corrected for,
    	and if the short-range correlations are dominated by Bose-Einstein
    	correlations, this two-particle Bose-Einstein correlation function
    	is related to the Fourier-transformed source distribution.
	For clarity, let us consider the case of a one-dimensional,
	factorized coordinate and momentum space distribution,
\begin{equation}
	S(x,k) =  f(x) \, g(k).
\end{equation}
	In this case~\cite{Csorgo:2003uv,Csorgo:2004ch},
	the Bose-Einstein correlation function is 
\begin{equation}
	C_2(k_1,k_2) = 1 + |\tilde f(q)|^2,
\end{equation}
	where the Fourier transformed source density (often referred to as the
	{\it characteristic function}) and the relative momentum are defined as
\begin{equation}
	\tilde f(q)  =  
		\int  \mbox{\rm d}x \, \exp(i q x) \,f(x),\qquad\quad \label{e:char}
	q  =  k_1 - k_2 .\label{e:fourier}
\end{equation}

	Let us focus on the property of particle emission from QCD jets, that the 
	fractal defining the particle emission is infrared stable:
	adding one more, very soft gluon does not change the
	resulting source distributions. Thus the  source of particles is stable
	for convolution. The Bose-Einstein correlation functions for such particle 
	emitting sources were evaluated recently by three of us,
	which we summarize in this section following refs.~\cite{Csorgo:2003uv} and ~\cite{Csorgo:2004ch}.

	For the case of the jets decaying to jets to jets and so on, the final position
	of a particle emission is given by a large number of position shifts,
	hence the distribution of the final position $x$ is obtained as  a convolution,
\be
        x = \sum_{i=1}^n x_i, \qquad\qquad
	        f(x) = \int \Pi_{i=1}^n \d x_i\, \Pi_{j=1}^n f_j(x_j)\,
            \delta(x - \sum_{k=1}^n x_k ).
\ee
	Various forms of the Central Limit Theorem state,
	that under certain conditions, the distribution of the sum of a
	large number of random variables converges (for $n \rightarrow \infty$) 
	to a limit distribution.  
	In case of ``normal" elementary processes, that have finite means
	and variances, the limit distribution of their sum is a Gaussian.
	Stable distributions are precisely those limit distributions
	that can occur in Generalized Central Limit theorems. Their study
	was begun by the French mathematician Paul L\'evy in the 1920's.
	The stable distributions are frequently given in terms of
	their characteristic functions, as the Fourier transform of
	a convolution is a product of the Fourier-transforms,
    \be
            \tilde f(q) = \prod_{i=1}^n \tilde f_i(q)
    \ee
	and limit distributions appear when the convolution
	of one more elementary process does not change the
	form of the limit distribution,  but it results only in
	a modification of its location and scale parameters.
	The characteristic function of univariate and symmetric stable distributions is
\be
	\tilde f(q)=\exp\left( i q \delta -|\gamma q|^\alpha\right), \label{e:fqs}
\ee
	where the support of the density function $f(x)$ is $(-\infty,\infty)$.
	Deep mathematical results imply that the index of stability, $\alpha$,
	satisfies the inequality $0 < \alpha \le 2$, 
	so that the source distribution be always positive.  
	These L\'evy distributions are indeed stable for convolutions,
    	in the following sense:
\bea
	\tilde f_i(q) & = & \exp\left( i q \delta_i -|\gamma_i q|^\alpha\right), \qquad
	\prod_{i=1}^n \tilde f_i(q) \, =  \,
	\exp\left( i q \delta -|\gamma q|^\alpha\right) ,\\
	\gamma^\alpha & = &  \sum_{i=1}^n \gamma_i^\alpha,
	\quad\qquad\qquad\qquad
	\delta \, = \, \sum_{i=1}^n \delta_i. \label{e:gami}
\eea

	Thus the Bose-Einstein correlation functions for univariate, 
	symmetric stable distributions (after a core-halo correction, and a re-scaling) read as
\begin{equation}
	C(q;\alpha) = 1 + \lambda \exp\left(-|q R|^\alpha\right).  \label{e:BEC-Levy}
\end{equation}
	For the special value of $\alpha = 2$ we obtain the well known Gaussian case.
	Refs.~\cite{Csorgo:2003uv} and ~\cite{Csorgo:2004ch} discuss further examples and details
    	and generalize these results to 
    	three dimensional, hydrodynamically expanding, core-halo type sources as
\begin{equation}
	C(q,K;\alpha) = 1 +\lambda(K)
       		\exp\left(-|\sum_{i,j} q_i R_{ij}^2(K) q_j |^{\alpha/2}\right).  
		\label{e:3dBEC-Levy}
\end{equation}

\begin{figure}[!thb]
\begin{center}
\includegraphics[angle=0,width=0.45\textwidth]{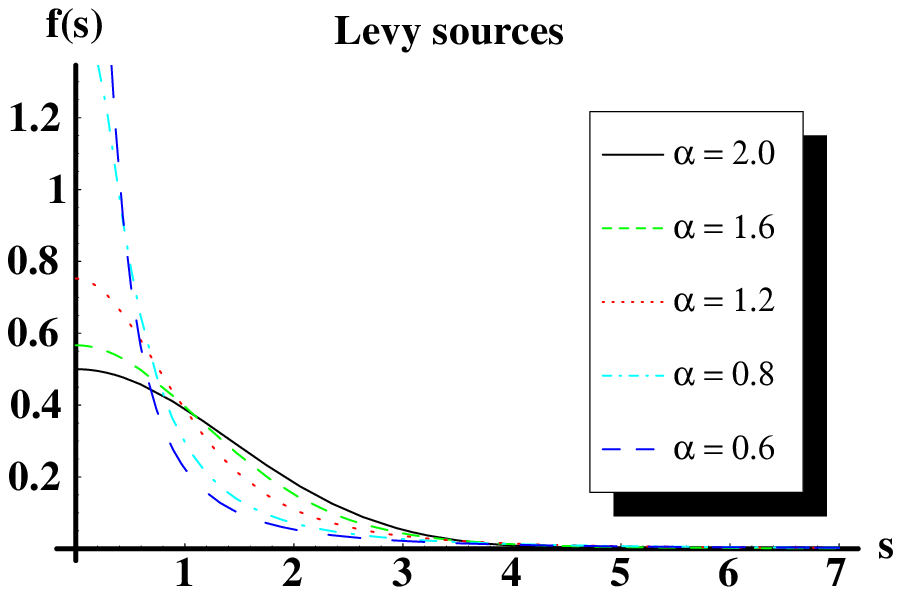}
\includegraphics[angle=0,width=0.45\textwidth]{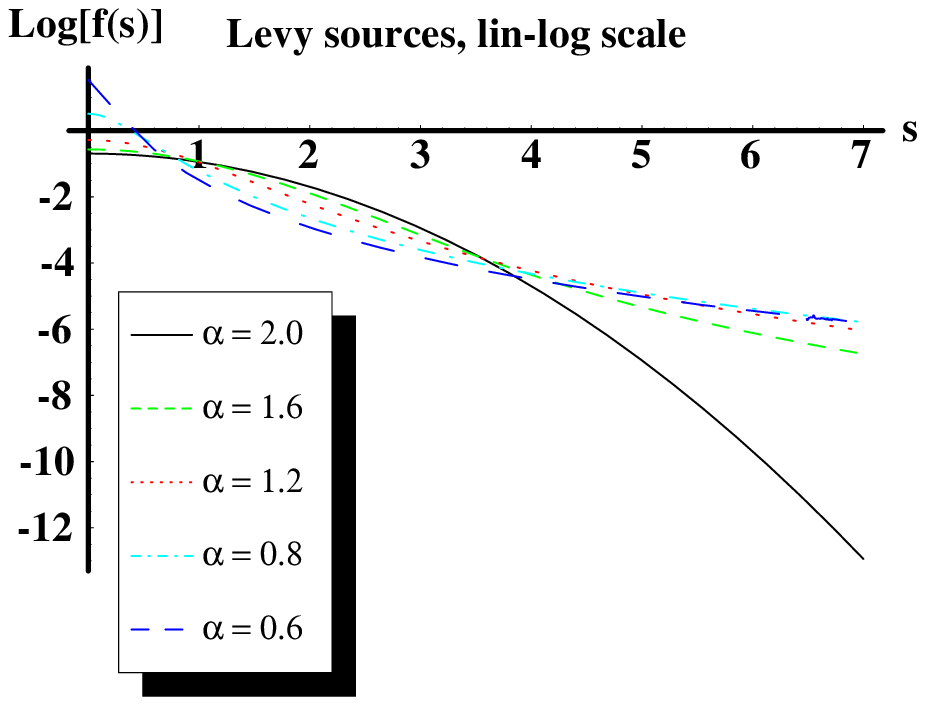}\\
\includegraphics[angle=0,width=0.45\textwidth,height=0.3\textwidth]{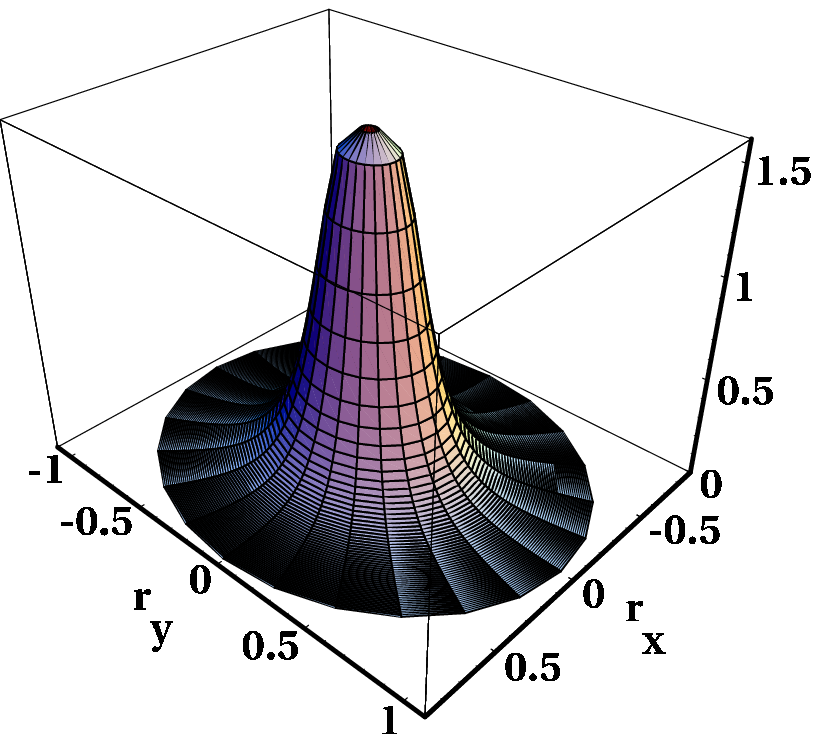} 
\includegraphics[angle=0,width=0.45\textwidth,height=0.3\textwidth]{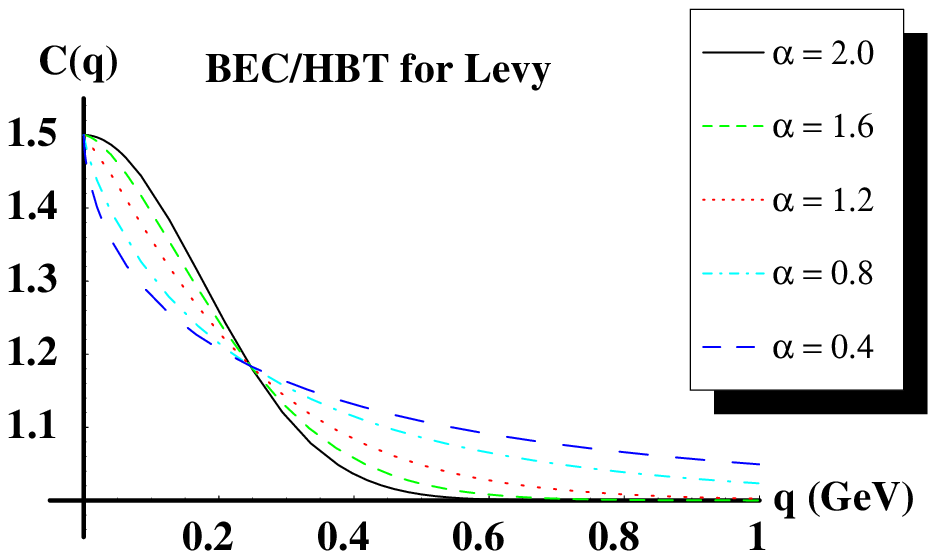} 
\\
\vspace{-0.5cm}
\end{center}
\caption[*]{
\label{fig:1dlevy-lin}{\small (a - top left) 
Source functions for univariate symmetric L\'evy laws, as 
a function of the dimensionless variable $s = r/R $, on a linear-linear scale,
for various values of the L\'evy index of stability, $\alpha$.
(b - top right) Same as (a), but on a linear-logarithmic plot.  Note the power-law
tails  that decay in case of  $ \alpha < 2$ 
for large values of $s$ as $f_\alpha(s) \propto s^{-1-\alpha}$.
(c - bottom left) Source function for two-dimensional
symmetric L\'evy laws, for $\alpha = 0.6$.
(d - bottom right) Bose-Einstein correlation (or HBT) correlation functions 
for univariate symmetric L\'evy laws, for a fixed  scale parameter of $R = 0.8$ fm 
and various values of the L\'evy index of stability, $\alpha$. 
} 
}
\end{figure}

	Figure~\ref{fig:1dlevy-lin} (a)  
	shows univariate symmetric L\'evy  source distributions.
	Figure~\ref{fig:1dlevy-lin} (b)  
	illustrates, that values of the index of stability $\alpha$ are related
    	to the tails of these distributions. If  $ \alpha < 2$,
    	for large values of $s=r/R$ the L\'evy sources decay 
       	as $ f_\alpha(s) \propto s^{-1-\alpha} $.
	Figure~\ref{fig:1dlevy-lin} (c)  
	shows a two-dimensional symmetric L\'evy stable source.
    	Bose-Einstein correlation functions for L\'evy stable source distributions are shown
	in  Figure~\ref{fig:1dlevy-lin} (d)  
       	for various values of the index of stability $\alpha$, and
	for a constant value of  the radius parameter $R$.
	These Bose-Einstein correlation functions are sensitive 
	to the value of $\alpha$ not only in  the small  $Q < \hbar/R$ region,
   	but are also in the ``large" relative momentum region of $ Q > \hbar/R$.
   	Thus these correlations are sensitive to the structure of the 
	particle emission in the region which is shaped by the jets. 

	A  random walk, where  the length of the steps is given by a L\'evy distribution,
        and the direction of the steps is random, corresponds to a fractal curve,
	in physical terms it can be interpreted as the path of a test particle
	performing a generalized Brownian motion. This motion is referred to
	as anomalous diffusion and the probability that the test particle diffuses
	to distances $r$ greater than a certain value of $|s| $ is given by
	$P(r > |s|)  \propto |s|^{-\alpha}$. This relation is valid for anomalous
	diffusion not only  in one, but also in  two and three dimensions.
	Thus the L\'evy index of stability $\alpha$ is the fractal dimension
	of the trajectory of the corresponding anomalous diffusion~\cite{Seshadri82ab}.
	When we apply this result to QCD, there are two key considerations.
       
       	First, if gluon radiation is neglected, the $q\overline{q}$ system
	hadronizes as a 1+1 dimensional hadronic string, which has no fractal
	structure. If the gluon emission is switched on, the emission of gluon $n$ from 
       one of the $n$ dipoles 	corresponds to a step of an anomalous diffusion 
	in the plane transverse to the given  dipole. Hence the anomalous 
	dimension of QCD equals to the L\'evy index of stability 
	of this anomalous diffusion,	
\be
	\sqrt{\frac{3 \alpha_s}{2\pi}} = \alpha_{\mbox{\tiny\rm L\'evy}}  . 
\ee	
	Second, data on Bose-Einstein correlations are often determined
       in terms of the invariant momentum difference $Q_{\rm inv} = \sqrt{-(p_1-p_2)^2} $.
	Bose-Einstein correlation functions that depend on this invariant momentum
       difference can be obtained within the framework of the so-called 
	$\tau$-model.       This model assumes a broad proper-time distribution,
	$H(\tau)$ and very strong correlations
	between coordinate and momentum in all directions, $x^\mu/\tau \propto p^\mu/m_{(t)}$.
	Hence $(x_1 - x_2) (p_1 - p_2) \propto \tau Q_{\rm inv}^2$,
	see refs.~\cite{cstjz,csorgo-alushta} for details.
	In this case, the Bose-Einstein correlation function measures the
	Fourier-transformed proper-time distribution $\tilde H$ in the following,
	unusual manner:
\be
	C_2(Q_{\rm inv}) \simeq 
	1 + \lambda \, \mbox{\it Re}\, \tilde H^2\left(\frac{Q_{\rm inv}^2}{2  m_{(t)}}\right),
\ee
	where $m_{(t)} $ stands for the (transverse) mass of the pair
	for (two)- or more jet events.	
	From this relation it follows, that $\alpha_{\mbox{\tiny\rm BEC}} = 
	2 \alpha_{\mbox{\tiny\rm L\'evy}}$.
	Thus we find the following relationship between the
	strong coupling constant and the exponent of an invariant relative momentum
        dependent Bose-Einstein correlation function:
\be
	\alpha_s = \frac{\pi}{6} \alpha^2_{\mbox{\tiny\rm BEC}}.\label{e:QCDa}
\ee

\section{\label{s:data} Application to NA22 and UA1 data}

	In ref.~\cite{Csorgo:2003uv} three of us have fitted the 
	NA22~\cite{na22-d2s} and the UA1 data~\cite{ua1-d2s}
	on two-particle Bose-Einstein correlation functions
	with the L\'evy stable form of eq. ~(\ref{e:BEC-Levy}).
	The results were summarized in Table 1 of that paper. 
	Here we re-interpret the exponent of this fit with the help
	of eq.~(\ref{e:QCDa}) and extract $\alpha_s$,
	the coupling constant of QCD, as given by Table 1 of this paper.

\begin{table}[htb]
\begin{center}
\begin{tabular}{|l|rl|rl|}
\hline
                 & \multicolumn{2}{c}{NA22 }
                 & \multicolumn{2}{|c|}{UA1}\\ \cline{2-5}
Parameter        & Value          & Error & Value & Error \\ \hline
$\alpha_{\mbox{\tiny\rm BEC}}$ from ref.~\cite{Csorgo:2003uv}         
                 & 0.67   &$\pm$ 0.07 
		 & 0.49   &$\pm$ 0.01
    		\\ \hline 
$\alpha_s$ from eq. (\ref{e:QCDa})  	  
    		 & 0.24   &$\pm$ 0.05
                 & 0.13   &$\pm$ 0.01   
		 \\ \hline
\end{tabular}
\end{center}
\caption
{
Converting the best L\'evy fits to UA1 and NA22 two-particle correlations 
    using the $\tau$-model converted with  eq.(\ref{e:QCDa}) to  
    values for the running QCD coupling constant $\alpha_s$.
}
\label{tab:results}
\end{table}

\section{Summary, conclusions}
Using the picture of strongly correlated coordinate and momentum space
distributions, we determined the (running of the)  strong coupling constant 
from NA22 and UA1 two-pion Bose-Einstein correlation measurements.

\section*{Acknowledgments}
T. Cs\"org\H{o} would like to thank Bill Gary and his team for a great meeting 
and to W. Kittel for inspiring discussions.
This work has been partially supported 
by  the Hungarian OTKA grant T038406, the Hungarian - US MTA - OTKA - NSF grant INT0089462 
and the Hungarian - Ukrainian - US NATO PST.CLG.980086 grant.


\begin{thebibliography}{99}
\bibitem{Bialas:1985jb}
	A.~Bialas and R.~Peschanski,
    	Nucl.\ Phys.\ B {\bf 273} (1986) 703.


\bibitem{Bialas:1990gt}
	A.~Bialas,
    	Nucl.\ Phys.\ A {\bf 525} (1991) 345.  

\bibitem{Dahlqvist:1989yc}
	P.~Dahlqvist, B.~Andersson and G.~Gustafson,
	Nucl.\ Phys.\ B {\bf 328} (1989) 76.

\bibitem{Gustafson:1990qi}
	G.~Gustafson and A.~Nilsson,
	Nucl.\ Phys.\ B {\bf 355} (1991) 106.

\bibitem{Gustafson:1990qk}
	G.~Gustafson, 
	{\it Proc Int. Workshop on Correlations and Multiparticle Production, 
	   Marburg, West Germany, May 14-16, 1990} 
	   (World Scientific, Singapore, 1991, eds. M. Pl\"umer, S. Raha and R. M. Weiner)

\bibitem{Gustafson:1991ru}
	G.~Gustafson and A.~Nilsson,
    	Z.\ Phys.\ C {\bf 52} (1991) 533.

\bibitem{Gustafson:1992uh}
	G.~Gustafson,
    	Nucl.\ Phys.\ B {\bf 392}, 251 (1993).

\bibitem{Andersson:1995jv}
	B.~Andersson, G.~Gustafson, J.~Samuelsson,
    	Nucl.\ Phys.\ B {\bf 463} (1996) 217.

\bibitem{Dokshitzer:1992df}
	Y.~L.~Dokshitzer and I.~M.~Dremin,
    	Nucl.\ Phys.\ B {\bf 402} (1993) 139.

\bibitem{Ochs:1992tg}
	W.~Ochs and J.~Wosiek,
    	Phys.\ Lett.\ B {\bf 305} (1993) 144.

\bibitem{Ochs:1994rt}
	W.~Ochs and J.~Wosiek,
    	Z.\ Phys.\ C {\bf 68} (1995) 269. 


\bibitem{DeWolf:1995pc}
	E.~A.~De Wolf, I.~M.~Dremin and W.~Kittel,
    	Phys.\ Rept.\  {\bf 270} (1996) 1

\bibitem{Bialas:1992ca}
    A.~Bialas,
    Acta Phys.\ Polon.\ B {\bf 23} (1992) 561.


\bibitem{Csorgo:2003uv}
    T.~Cs\"org\H{o}, S.~Hegyi and W.~A.~Zajc,
    Eur.\ Phys.\ J.\ C {\bf 36}, 67 (2004)

\bibitem{Csorgo:2004ch}
    T.~Cs\"org\H{o}, S.~Hegyi and W.~A.~Zajc,
    arXiv:nucl-th/0402035.

\bibitem{Utyuzh:1999zg}
	O.~V.~Utyuzh, G.~Wilk and Z.~Wlodarczyk,
    Phys.\ Rev.\ D {\bf 61}, 034007 (2000).

\bibitem{Seshadri82ab}
	V. Seshardi, B. J. West,
	Proc. Nat. Acad. Sci. USA {\bf 79} pp. 4501 - 4505 (1982)

\bibitem{cstjz} 
	T. Cs\"org\H{o} and J. Zim\'anyi, 
	Nucl. Phys. A {\bf 517} (1990) 588

\bibitem{csorgo-alushta}
	T. Cs\"org\H{o}, in
	Proc. XXXII Int. Symp. Multipart. Dynamics [hep-ph/0301164]

\bibitem{na22-d2s}
	N.~M.~Agababian {\it et al.}  [EHS/NA22 Coll.],
    Z.\ Phys.\ C {\bf 59}, 405 (1993).



\bibitem{ua1-d2s}
    N.~Neumeister {\it et al.}  [UA1-Min. Bias-Coll.],
    Z.\ Phys.\ C {\bf 60}, 633 (1993).


\end{thebibliography}
\end{document}